\documentclass[a4paper,10pt,reqno]{amsart}
\usepackage[utf8]{inputenc}
\usepackage[%
    hyperref,backend=biber, doi=false,url=false,
    sorting=none,style=numeric-comp,backref=true, maxbibnames=4]{biblatex} 

\usepackage{amsmath}
\usepackage{amssymb}
\usepackage{bbm}
\usepackage{color}

\usepackage{graphicx}
\usepackage{epstopdf}
\usepackage{hyperref}
\hypersetup{colorlinks=true,linktoc=page,linkcolor=blue,citecolor=red,urlcolor=cyan}

\newcommand{\colblack}[1]{#1}

\bibliography{bibliografia}

\usepackage[english]{babel}

\newcommand{\paras}[1][x]{\mathbf{#1}^\parallel}
\newcommand{\para}[1][x]{{#1}^{\parallel}}
\newcommand{\perpe}[1][x]{#1^D}
\newcommand{\dd}{\Delta_{\delta}}
\newcommand{\posi}{L}
\newcommand{\ampli}{\zeta}
\newcommand{\kernel}{\gamma}
\newcommand{\itkernelzero}{\Gamma}
\newcommand{\itkernel}{\itkernelzero_2}
\newcommand{\itkernelc}{\bar{\Gamma}_2}
\newcommand{\Kd}{K_{\delta}}
\newcommand{\ff}{F}
\newcommand{\A}{F_1}
\newcommand{\B}{F_2}
\newcommand{\C}{F_3}
\newcommand{\aaa}{a}

\title[Effective action for delta potentials]{Effective action for delta potentials: spacetime-dependent inhomogeneities and Casimir self-energy}
\author{S.~A.~Franchino-Vi\~nas$^{1,2}$}

\address{$^1$ Departamento de F\'isica, Facultad de Ciencias Exactas
Universidad Nacional de La Plata, C.C.\ 67 (1900), La Plata, Argentina.}

\address{$^2$ Institut für Theoretische Physik, Universität Heidelberg, D-69120 Heidelberg, Germany.}
\email{safranchino@fisica.unlp.edu.ar}

\author{F.~D.~Mazzitelli$^{3,4}$}

\address{$^3$ Centro At\'omico Bariloche,  CONICET,
Comisi\'on Nacional de Energ\'\i a At\'omica, R8402AGP Bariloche, Argentina}

\address{$^4$
Instituto Balseiro, Universidad Nacional de Cuyo, R8402AGP Bariloche, Argentina. }

\email{ fdmazzi@cab.cnea.gov.ar}

\begin{document}

\begin{abstract}
We study the vacuum fluctuations of a quantum scalar field in the presence of a thin and inhomogeneous
flat mirror, modeled with a delta potential. Using Heat-Kernel techniques, we evaluate the Euclidean effective action perturbatively in the inhomogeneities (nonperturbatively in the constant background). We show that the divergences can be absorbed into a local counterterm, and that the remaining finite part is in general a nonlocal functional of the inhomogeneities, which we compute explicitly for massless fields in $D=4$ dimensions. For time-independent inhomogeneities, the effective action gives the Casimir self-energy for a partially transmitting mirror. For time-dependent inhomogeneities, the Wick-rotated effective action
gives the probability of particle creation due to the dynamical Casimir effect.
\end{abstract}

\maketitle

\section{Introduction}\label{sec:intro}
Quantum fields on nontrivial backgrounds or in the presence of boundary conditions are of interest in many branches of physics. Vacuum fluctuations of the quantum fields produce interesting physical phenomena like Casimir forces and particle creation by moving mirrors or by time-dependent gravitational fields. They can also serve as seeds for structure formation in inflationary models. 

When analyzing the static Casimir effect \cite{Milton:2001yy, Bordag:2009zz} beyond the
perfect conductor limit, the physical properties 
of thin surfaces  can be described by delta potentials, that is, potentials that are proportional to  Dirac delta functions with support on those surfaces. For scalar fields, the analog of 
perfect conductor are Dirichlet or Neumann boundary conditions, 
that can be obtained from a delta potential in the limit in which the coefficient of the delta function (or its derivative)
tends to infinity. Otherwise, the delta potential models a thin semitransparent mirror. Here we will be concerned with inhomogeneous mirrors, 
on whose surfaces the physical properties vary from point to point and may also depend on time. 

Casimir forces for homogeneous delta potentials have been analyzed previously in the literature for different geometries and using different methods \cite{Milton:2004ya, Milton:2007gy, Vinas:2010ix, Munoz-Castaneda:2013yga}. The dynamical Casimir effect \cite{Dodonov:2010zza, Dalvit:2010ria, Nation:2011dka} for semitransparent mirrors has also been considered both for  moving
mirrors \cite{Frolov:1999bi, Fosco:2007nz, Haro:2008zza, Fosco:2011rm, Fosco:2017jjf} and for static mirrors with time-dependent properties
\cite{Crocce:2004jq, Silva:2011fq}. 

The situation for Casimir self-energies is more subtle. Although not relevant for the computation of forces
between different bodies, the vacuum energy, or more generally the vacuum expectation value of the energy-momentum tensor for the quantum fields,  couples to gravity through the semiclassical Einstein equations. It is also relevant in the calculation of the Casimir self-stresses on curved surfaces \cite{Parashar:2017sgo}.
As pointed out a long time ago  by Deutsch and Candelas \cite{Deutsch:1978sc}, the renormalized energy-momentum
tensor diverges near a perfect conductor, and the divergence
depends on the local geometry of the surface. This divergence implies that the total self-energy is also divergent, although there can be partial cancellations between divergences on both sides of the surface. Local and global divergences for homogeneous delta potentials has been discussed by Milton in Ref. \cite{Milton:2010qr}, while the divergent part of the effective action for inhomogeneous delta potentials
has been computed by Bordag and Vassilevich \cite{Bordag:2004rx} using Heat-Kernel (HK) techniques. More general situations have also been considered, in which the thin surface is replaced by a layer of finite width \cite{Graham:2002xq}. The layer can be modeled with a classical background field that interacts with the vacuum field. In this case, the self-energy can be made finite renormalizing the 
classical background.  This approach can be generalized for
quantum fields on a curved spacetime, taking into account the contribution of the vacuum
expectation value of the energy-momentum tensor to the 
semiclassical Einstein equations, and renormalizing the theory using
standard techniques \cite{Mazzitelli:2011st}. Different aspects of
the vacuum energy in the presence of soft walls have been discussed
in Refs. \cite{Milton:2011iy, Milton:2016sev}.

\colblack{Regarding the renormalization, it is customary in the Casimir context to renormalize by introducting classical fields, in such a way that in the large-mass limit the renormalized energy should vanish \cite{Bordag:1998vs,Bordag:1999sf,Kirsten:2001wz}.
Although this criterion is clearly not available for the massless case, the renormalization procedure can still be consistently applied. 
In the HK framework, it has been shown that there are two different scenarios, depending on whether a given coefficient in the small propertime Seeley-DeWitt expansion vanishes or not \cite{Bordag:1998vs}.
If it doesn't vanish then the procedure loses some predictivity because one should fix some constants by using ``experimental data". In any case, it should be clear that this does not imply an ambiguity in the process, but just a partial lost in predictivity.
}

In this paper, we will work within the thin surface idealization, 
and consider an inhomogeneous delta potential on a single flat surface. We will go beyond
the above mentioned works on singular potentials by computing the effective action in an expansion in powers of the inhomogeneities, paying attention not only to the divergent part, but also to its 
finite contribution. 
The effective action will be a nonlocal functional 
of the inhomogeneities. Moreover, when the properties 
of the mirror depend on time, the time-dependent boundary conditions will produce particle creation. This is
a particular realization of the dynamical Casimir effect,
in which the mirror is static but its physical properties depend on time. 
We will compute the vacuum persistence amplitude from the imaginary part of the Lorentzian effective action. 
Several of our results for singular potentials have their counterpart in semiclassical gravity, as we will point out along the paper. 

From a technical point of view, we will use HK techniques. We will first consider a massive scalar field in an
Euclidean space of $D$ dimensions. As described in Sec. \ref{sec:HK}, 
the HK and the Euclidean effective action can be evaluated exactly for homogeneous delta potentials, and perturbatively for small departures
from homegeneity. In Sec. \ref{sec:D4m0} we will analyze in detail the massless case in $D=4$ dimensions. We will discuss the 
divergences of the effective action and analyze its finite part in  two opposite limiting situations:  smooth and rapidly varying inhomogeneities. For time-independent inhomogeneities, explicit evaluations of the self-energy of the plate are described in Sec. \ref{sec:examples}.  We will then consider the case of time-dependent inhomogeneities, and its relation with the dynamical Casimir effect. As shown in Sec. \ref{sec:DCE}, the Lorentzian (or in-out) effective action can be obtained from the Euclidean effective action through a Wick rotation. The imaginary part of the effective action, which is a finite quantity, gives the vacuum persistence amplitude for this problem. We will obtain a general expression, and then discuss some particular examples. Sec. \ref{sec:conclus} contains the conclusions of our work. The Appendixes describe some details of the calculations, as well as an alternative derivation of the Casimir self-energy using an approach based on the Gel'fand-Yaglom theorem \cite{Fosco:2019lmw}.

\section{Effective action in presence of a delta potential: the Heat-Kernel approach}\label{sec:HK}
As said above, our goal is to study the quantum properties of the vacuum for a field in presence of a singular potential $V$,
namely a Dirac delta function whose support lies on a hypersurface of codimension one. 
One effective way to perform such a study, is by using spectral functions. Let us briefly review how  this can be accomplished.

Consider then a massive real scalar field $\phi$ defined on a $D$-dimensional Euclidean flat space and in presence of an external potential $V(x)$; 
accordingly, its action reads 
\begin{align}\label{eq:action_phi}
 S_{\phi}&=\frac{1}{2}\int d^Dx\, \phi(x)\left(-\partial^2+ m^2+V(x)\right)\phi(x).
\end{align}
As customary, one can perturbatively integrate the quantum field in order to obtain the effective action, 
from which it is easier to study the desired properties.
For the action \eqref{eq:action_phi}, it is straightforward to show that the 1-loop contribution to the effective action,
which is actually  the only quantum correction in this simple case,
%\footnote{Possibly up to terms coming from the renormalization process.}\href{https://arxiv.org/pdf/2006.03822.pdf}{este artículo y las referencias me habían hecho pensar que quizás estaba equivocado; el cálculo directo y otros argumentos dicen que no hay más correcciones},
 is given by
\begin{align}
 \Gamma_{\text{1-loop}}=\frac{1}{2}\text{Tr}\,\text{Log} \left(-\partial^2+ m^2+V(x) \right).
\end{align}
Alternatively, by using Schwinger's trick \cite{Schubert:2001he,Kirsten:2001wz} one may recast this expression into 
\begin{align}\label{eq:EA_1loop_Schwinger}
 \Gamma_{\text{1-loop}}=-\frac{1}{2}\int_0^{\infty} \frac{dT}{T} \text{Tr}\, K_V,
\end{align}
where we have defined the HK of the operator of quantum fluctuations as
\begin{align}
 K_V:=e^{-T (-\partial^2+ m^2+V(x))}.
\end{align}
We have thus reduced our problem to the determination of a relevant HK, task to which we will devote the following subsections.

\subsection{The exact Heat-Kernel for a homogeneous background}
As a warm-up, let us begin by considering the operator 
\begin{align}\label{eq:dd}
 \dd := -\partial^2_x + \ampli\, \delta(x- \posi),\quad x\in\mathbb{R},
\end{align}
for a constant $\ampli>0$ (as long as the contrary is not stated, this assumption will be considered throughout the rest of the article). 
Up to our knowledge, its exact HK has been obtained for the first time in 
\cite{Bauch_1985} by considering path integral and Laplace-transform techniques; later, it was independently rederived in \cite{Gaveau_1986}, 
in a similar way (employing methods of Brownian motion and Laplace transforms). 
Here we will provide a new derivation of an exact expression for its HK, which involves solving an integral equation
derived from a path integral representation.

Recall that the problem of finding the HK of $\dd$ is equivalent to the
determination of the propagator of a particle in a first-quantization realm and subject to 
a potential 
which is a Dirac delta, to wit
\begin{align}\label{eq:worldline_d1}
 \Kd(x,y;T):%&=e^{-T \dd}(x,y)\\
 &=\int_{\substack{q(0)=x}}^{q(T)=y} \mathcal{D}q(t) e^{- S_{\delta}},
\end{align}
where 
\begin{align}
 S_{\delta}=\int_0^T dt \, \left[\frac{\dot{q}^2(t)}{4}+\ampli\,\delta(q(t)- \posi)\right].
\end{align}
Notice, of course, that this corresponds to a ficticious particle that evolves in time $t$, which is ficticious as well. 
This implementation of path integrals in the computation of HKs and effective actions
is usually called Worldline Formalism or String-inspired Formalism in the literature \cite{Schubert:2001he,Edwards:2019eby}.
In relation to Dirac delta potentials, a first-order perturbative computation of the Casimir energy between two semitransparent layers has 
been computed in \cite{Vinas:2010ix}, while other recent applications include computations in QED \cite{Nicasio:2020awj} and noncommutative Quantum Field Theory \cite{Franchino-Vinas:2018jcs}. 

Coming back to Eq.\eqref{eq:worldline_d1}, one can formally perform the usual expansion in powers of $\ampli$, which after choosing a 
convenient ordering in the intermediate times $t_i$ leads to
\begin{align}\label{eq:K_series}
 \begin{split}
\Kd(x,y;T)&= \sum_{n=0}^{\infty} (-\ampli)^n\int_0^T dt_n \int_0^{t_n} dt_{n-1}\cdots \int_0^{t_2} dt_1  K_0(y,\posi; T-t_n) \\
 &\hspace{1.5cm}\times  K_0(\posi,\posi; t_n-t_{n-1}) \cdots K_0(\posi,\posi; t_2-t_1)  K_0(\posi, x; t_1),
 \end{split}
\end{align}
where $K_0$, the free HK,  has the expression
\begin{align}
 K_0(x,y;T):=\frac{e^{-\frac{(x-y)^2}{4T}}}{\sqrt{4\pi T}}.
\end{align}
From Eq. \eqref{eq:K_series}, one can show that the HK satisfies an integral equation whose kernel is nothing but the HK of the free particle \cite{Bordag:1999ed}. 
This last fact obstructs the obtention of the solution via iterated kernels and one then needs to recur to Laplace-transform techniques. 

However, one may also derive from Eq.\eqref{eq:K_series}  a simpler integral equation. In order to do so, 
we perform a change of variables $s_1:=t_1\, s_2:=t_2-t_1,\, s_3:=t_3-t_2,$ etc., and obtain
\begin{align}\label{eq:K_integral_eq}
 \Kd(x,y;T)
 &= K_0(x,y;T) - \ampli \int_0^T ds\, K_0(x,\posi;s) f(s),
 \end{align}
where the function $f(s)$ is defined as
\begin{align}
 \begin{split}
f(s_1):&= K_0(y,L;T-s_1)- \ampli \int_0^{T-s_1}  \frac{ds_2}{\sqrt{4\pi s_2}}\Big[K_0(y,L;T-s_1-s_2)\\
&\hspace{1cm} - \ampli \int_0^{T-(s_1+s_2)} \frac{ds_3}{\sqrt{4\pi s_{3}}} \Big( K_0(y,L;T-s_1-s_2-s_3)+ \ldots \Big)\Big].  
 \end{split}
\end{align}
Now it should be clear that $f(s)$ satisfies an integral equation of Volterra type involving a rather simple kernel,
\begin{align}
 f(s_1)&=K_0(y,\ampli;T-s_1)- \frac{\ampli}{\sqrt{4\pi}} \int_{0}^{T} ds_2 \frac{\Theta(s_2-s_1)}{\sqrt{ (s_2-s_1)}} f(s_2),
\end{align}
whose solution can be straightforwardly obtained by considering the method of the iterated kernel. Following this path we get
\begin{align}
 f(s)&=\int_0^T ds_1 \,\itkernel(s_1-s) K_0(y,\posi,T-s_1),
 \end{align}
where $\itkernel$, the series of iterated kernels, can be resummed as
\begin{align}\label{eq:iterated_kernel}
 \begin{split}\itkernel(s_2-s_1):&= \delta(s_1-s_2)-\frac{\ampli}{4} \Theta(s_2-s_1) \\
 &\hspace{2cm}\times \left(\frac{2}{\sqrt{\pi(s_2-s_1)}}-\ampli e^{\frac{\ampli^2}{4}(s_2-s_1)}   \text{erfc}\left(\frac{\ampli}{2} \sqrt{s_2-s_1}\right)\right).
\end{split}
\end{align}
We have additionally introduced the complementary error function
 \begin{align}
 \text{erfc}(x):=\frac{2}{\sqrt{\pi}}\int_x^{\infty} e^{-t^2}dt.
\end{align}
At this point, a direct replacement of the obtained $f(s)$ in Eq.\eqref{eq:K_integral_eq}, yields the following expression for the HK:
\begin{align}\label{eq:HK_1d}
 \begin{split}\Kd(x,y;T)
 &= K_0(x,y;T)\\
 &\hspace{0.5cm}- \ampli \int_0^T\int_0^T ds_1\,ds_2\, K_0(x,\posi;s_1) \,\itkernel(s_1-s_2) K_0(y,\posi,T-s_2).
 \end{split}
 \end{align}
The proofs of the fact that this expression is equivalent to the one obtained in \cite{Bauch_1985} and that it explicitly 
satisfies the so-called convolution property of propagators,
\begin{align}\label{eq:HK_convolution}
 \Kd(x,y;S)=\int_{\mathbb{R}} dz\, \Kd(x,z;S-T) \Kd(z,y;T),
\end{align}
are left to Appendixes \ref{app:Bauch} and \ref{app:convolution} respectively.

Besides these local expressions, one may also consider the trace of the HK. A direct computation gives
\begin{align}\label{eq:HK_1d_integrated}
\begin{split}
\Kd(T):&=\int dx\, \Kd(x,x;T)\\
&=\frac{\int dx}{\sqrt{4\pi T}}+ \frac{1}{2}\left(e^{\frac{\ampli^2T}{4}} \text{erfc}\left(\frac{\sqrt{T}\ampli}{2}\right)-1\right).
\end{split}
\end{align}
As customary, the leading term in a small-propertime ($T$) expansion involves the volume of the manifold, 
which in  Eq.\eqref{eq:HK_1d_integrated} has been expressed as a divergent integral over the whole real 
line\footnote{A formal treatment of this divergence involves the introduction of a smearing function \cite{Vassilevich:2003xt}.}.
Although Eq.\eqref{eq:HK_1d_integrated} agrees with the results obtained in \cite{Bordag:1999ed} 
for the first coefficients in a small propertime expansion, it disagrees with
the expression stated in \cite{Munoz-Castaneda:2013yga}. In effect, beyond the fact that they already include the mass contribution to the 
HK, we have an additional constant term ($-1/2$). 
The reason for this discrepancy seems to lie on the regularization chosen in \cite{Munoz-Castaneda:2013yga},
where the authors consider a large interval of length $L$, at which ends they impose periodic boundary conditions.
In any case, the importance of this constant factor can be seen both in the $\ampli\to 0$ and $\ampli\rightarrow \infty$ limits: 
in the first limit one recovers the trace of the free HK, while in the second one Eq.\eqref{eq:HK_1d_integrated} reproduces the trace of the Dirichlet propagator, as expected.

\subsection{Perturbative computations in an inhomogeneous background}
Let us now generalize the operator in Eq. $\eqref{eq:dd}$ to a flat $D$-dimensional Euclidean space.
To simplify the notation and without loss of generality, 
we divide the coordinates as $x=(\para,\perpe)$, 
where $\para$ are the $D-1$ coordinates parallel to the plate and $\perpe$ the perpendicular one. 
In addition, we will allow a dependence of the delta's coupling on the parallel coordinates
through inhomogeneities that will be called $\eta(\para)$. 
Notice that a possible dependence on the Euclidean time is included as well.
The operator of our interest is thus
\begin{align}
 \dd^{(D)} = -\partial^2 + \left(\ampli+\eta(\para)\right) \delta(\perpe- \posi).
\end{align}
We will not focus on the precise field theory that gives origin to this operator. Just to mention an example, it could be the case that $\ampli$ were a coupling constant and $\eta$ were a field. However, we could also think of $\ampli+\eta(\para)$ as the vacuum expectation value of a quantum field that interacts with the field $\phi$. We will come back to this issue later on, in Sec. \ref{sec:D4m0}, where we analyze the emergency of divergencies in the massless  model for $D=4$.

In any case, as done in the previous subsection, its HK can be interpreted in the Worldline Formalism as
\begin{align}\label{eq:HK_D}
 K(x,y;T) &=\int_{\substack{q(0)=x}}^{q(T)=y} \mathcal{D}q(t) e^{- S^{(D)}},
\end{align}
if one introduces the appropriate first-quantization action, namely
\begin{align}
 S^{(D)}=\int_0^T dt\, \left[ \frac{\dot{q}^2(t)}{4}+\ampli\left(q_{\parallel}(t)\right) \delta(q_{D}(t)- \posi)\right].
\end{align}

Expression \eqref{eq:HK_D} has the advantage that it is specially well suited for perturbative computations. 
In fact, if we consider small inhomogeneities, the expansion of the HK up to quadratic order in $\eta$ reads
\begin{align}\label{eq:HK_D_expansion}
 \begin{split} K(x,y;T)
 &=\int_{\substack{x(0)=x}}^{x(T)=y} \mathcal{D}x(t)  e^{- \int_0^T dt\, \left[\frac{1}{4}\dot{x}^2(t)- \ampli \delta(\perpe(t)- \posi)\right]} \left[1+ \int_0^T ds_{1} \eta(\para_1)  \delta(\perpe_1- \posi)\right.\\
 &\hspace{2cm}\left.+\frac{1}{2} \int_0^T ds_{1}ds_2 \eta(\para_1) \eta(\para_2)  \delta(\perpe_1- \posi)  \delta(\perpe_2- \posi) +\cdots\right],
\end{split}
\end{align}
where the subscripts in the coordinates are devised to describe their dependence on the $s$ parameters, i.e. $x_i:=x(s_i)$.
Two comments are now in order. First, the $\eta$-independent term in formula \eqref{eq:HK_D_expansion} can be computed exactly,
given that it factorizes into the product of the HK of the Laplacian in $D-1$ dimensions, times the HK of the operator $\dd$ previously studied (from this zeroth-order term one can compute the effective
action for the homogeneous case, $\Gamma^{(0)}$).  
Second, the linear term will be considered to vanish by assuming that the mean of the inhomogeneities vanishes.

Having said so, we will focus on the term which is quadratic in $\eta$ in Eq.\eqref{eq:HK_D_expansion}, 
and compute its contribution to the trace of the HK, which will be called $K^{(2)}$. 
The computation is standard, albeit lengthy; we will consequently omit the details that lead us to
\begin{align}\label{eq:HK_D_quadratic}
K^{(2)}(T)%:&=\int d^Dx K^{(2)}(x,x;T)\\
:&=\int \frac{d\para[k]}{(2\pi)^{D-1}} \, \vert \tilde\eta(\para[k]) \vert^2 \kernel(\para[k],\ampli,T),
\end{align}
where we have introduced the Fourier transform of the inhomogeneities,
\begin{align}
 \eta(\para)=: \int \frac{d^{D-1}\para[k]}{(2\pi)^{D-1}}\,e^{i \para[k] \para }\tilde\eta(\para[k]),
\end{align}
as well as the kernel
\begin{align}\label{eq:kernel_gamma}
 \begin{split}
\kernel(\para[k],\ampli,T) :&= \frac{1 }{(4\pi )^{\frac{D-1}{2}}} \frac{1}{T^{\frac{D-5}{2}}}
 \int_0^1 ds (1-s) e^{ -T s \left(1-s\right)(\para[k])^2 }\\
 &\hspace{4cm}\times \Kd\left(\posi,\posi;T(1-s)\right) \Kd(\posi,\posi;Ts)  .
 \end{split}
\end{align}

By replacing these results into Eq.\eqref{eq:EA_1loop_Schwinger}, we obtain our master formula for the contribution to the effective action 
which is quadratic in $\eta$,
\iffalse
\begin{align}\label{eq:master}
\Gamma^{(2)}:= -\frac{1}{2}\int \frac{d\para[k]}{(2\pi)^{D-1}}  \, \vert \tilde\eta(\para[k])\vert^2 \int_0^{\infty}\frac{dT}{T} e^{-m^2 T}\kernel(\para[k],\ampli,T),
\end{align}
\fi
\begin{align}\label{eq:master}
\Gamma^{(2)}:= -\frac{1}{2}\int \frac{d\para[k]}{(2\pi)^{3}}  \, \vert \tilde\eta(\para[k])\vert^2 \ff(\para[k],\ampli,m^2),
\end{align}
written in terms of the form factor
\begin{align}\label{eq:form_factor}
 \ff(\para[k],\ampli,m^2):=\frac{1}{(2\pi)^{D-4}}\int_0^{\infty} \frac{dT}{T} e^{-m^2 T}\kernel(\para[k],\ampli,T).
\end{align}
Formula \eqref{eq:master} is valid for any dimension $D$ and for regular inhomogeneities $\eta$ that, as stated before, could also depend on the Euclidean time.
Notice also that it  is nonperturbative (or exact) in the (constant) coupling $\ampli$ and nonlocal, because of the form factor's $\para[k]$ dependence.

In the particular case $\ampli=0$, the form of the result \eqref{eq:master} would remind the reader the expression usually obtained when considering quantum fields
on weak nonsingular backgrounds \cite{Barvinsky:1987uw, Avramidi:1990je}. When $\ampli\neq 0$, the resummation implied in the form factor involves contributions coming from terms with any power
of the singular background field.

Coming back to expression \eqref{eq:master}, after a rescaling of $T$
we obtain
\begin{align}\label{eq:form_factor2}
 \ff(\para[k],\ampli,m^2):=\frac{1}{(2\pi)^{D-4}}\int_0^{\infty} \frac{d\tau}{\tau} e^{-\tau}\kernel(\para[k],\ampli,\tau/m^2),
\end{align}
which is suitable for an expansion in powers of $(\para[k])^2/m^2$. 
This would be the analog of the Schwinger-DeWitt expansion \cite{Barvinsky:1985an},
that in this case becomes a formal (local) series expansion in terms of derivatives
of $\eta(\para)$.  For massless fields, we expect in general a nonlocal effective action. 
Even if the presence of the additional scale $\ampli$ could make us think the contrary,
in the next section we will see that  the effective action does not admit
a derivative expansion, that is, the form factor cannot
be expanded in integer powers of $(\para[k])^2/\ampli^2$.

Unfortunately, a closed expression for the integrals involved in expression \eqref{eq:master} is not available to us for arbitrary dimensions.
Nevertheless, having in mind a dimensional regularization, we can analyze the region of convergence of the integral in the complex $D$-plane.
In order to perform such an analysis notice that, if $\ampli>0$, the HK at coincident points possesses the following asymptotic expansions:
\begin{align}\label{eq:kd_local_expansions}
 \Kd(\posi,\posi;T)\sim
 \begin{cases}
 \frac{1}{\sqrt{4\pi T}} -\frac{\ampli}{4}+\cdots,\quad &T\ll1
 \\
 \frac{1}{\sqrt{\pi}\ampli^2 } \frac{1}{T^{3/2}}-\frac{6}{\sqrt{\pi}\ampli^4} \frac{1}{T^{5/2}}+\cdots, \quad &T\gg1
\end{cases} .
\end{align}
This means that for small $T$, and disregarding a convergent integral in $s$, we have the power counting
\begin{align}
 \frac{1}{T} \kernel(\para[k],\ampli,T)\sim \frac{1}{T^{\frac{D-1}{2}}} ,\quad T\ll1,
\end{align}
so that we should impose $\operatorname{Re} D<3$ if we desire a convergent integral in $T$. 

On the remaining limit, namely for large $T$, if the field is massive there is no additional restriction.
Instead, the massless case is more subtle: we can split the $s$ integral into two,
one for $s\in(0,1/2)$ and the other for $s\in(1/2,1)$. 
Each of these integrals contains a propagator whose expansion for large $T$ provides an 
additional  factor $T^{-3/2}$, while not interfering in the convergence of the $s$ integral;
this yields a power counting 
\begin{align}
 \frac{1}{T} \kernel(\para[k],\ampli,T)\sim \frac{1}{T^{\frac{D+1}{2}}} ,\quad T\gg1.
\end{align}
In other words, one needs to impose the further condition $\operatorname{Re}D>1$ to guarantee the convergence of both the integrals in $T$ and $s$.

Of course one could made a rigorous proof of the preceding statements, 
for example considering Fubini's theorem and the bound \eqref{eq:erfc_bound2} from Appendix \ref{app:erfc}. Without entering into details, for a sufficiently regular $\eta(\para[k])$ the expression \eqref{eq:master} is well defined at least in the region
in which $1<\operatorname{Re} D<3$. In particular, in the physically relevant case $D=4$ a regularization should be provided.
This will be performed in the following section.

\section{A massless field in $D=4$}\label{sec:D4m0}
\subsection{Computation of the form factor and renormalization}
Consider now the situation of a massless field living in $D=4$. 
As explained before, the integrals involved in expression \eqref{eq:master} is in principle valid in the region $1<D<3$.
We will now see that its analytic continuation to $D=4$ shows a pole, forcing us to perform a renormalization.

To simplify the task, we will divide the form factor into three terms that we will define in the following\footnote{The region of convergence of each term alone can be proved to contain the
region $2<\operatorname{Re} D<3$ by employing the bound \eqref{eq:erfc_bound1} in App. \ref{app:erfc}.},
\begin{align}
  \ff(\para[k],\ampli) =\A+\B+\C.
\end{align}
The first one is made from the contribution of two free propagators and reads
\begin{align}
 \begin{split}
\A:&=\frac{1}{ 2^{2 D-3} \pi ^{\frac{3 D-7}{2}} }\int_0^{\infty} \frac{dT}{T^{\frac{D-3}{2}}}   \int_0^1 ds (1-s)  e^{ -s T \left(1-s\right)(\para[k])^2 } \frac{1}{T\sqrt{s(1-s)}}\\
 &=\frac{1}{2^{3 D-5} \pi ^{\frac{3 D}{2}-4} } \frac{\Gamma\left(\frac{3-D}{2}\right)\Gamma\left(\frac{D-2}{2}\right)}{\Gamma\left(\frac{D-1}{2}\right)}
 \vert \para[k]\vert^{D-3}.
 \end{split}
\end{align}
In the limit where $D\to 4$ we obtain
\begin{equation}
F_1=-\frac{ \vert\para[k]\vert}{32\pi^2}\, .
\end{equation}

The second one, comes from the sum of two contributions, each one coming from the product of one free propagator and one erfc function:
\begin{align}
 \begin{split}
\B:&=-\frac{\ampli }{2^{2 D-3} \pi ^{\frac{3 D-7}{2}}} \int_0^{\infty} \frac{dT}{T^{D/2-1}}  \int_0^1 \frac{ds }{\sqrt{s}}  e^{ -{s  \left(1-s\right) T(\para[k])^2 }{}} \int_0^{\infty} du e^{- \frac{\ampli}{2} u} \frac{e^{-\frac{u^2}{4T(1-s)}}}{\sqrt{ T (1-s)}}\\
 &=-\frac{\ampli^{5-D} }{2^{2 D-4} \pi ^{\frac{3 D-7}{2}} }\int_0^1 ds \int_0^{\infty} du  \int_0^{\infty} \frac{dT}{T^{(D-3)/2}}  \frac{\left(1-s\right)^{D/2-2}}{\sqrt{s}} e^{ - T(\aaa^{-2}s+ u+u^2)}.
\end{split}
\end{align}
The simplifications in the second line involve the successive rescalings $T\rightarrow \frac{T}{(1-s)\ampli^2}$, $u\rightarrow \frac{2}{\ampli}T u$, and the definition $\aaa=\frac{\left \vert \ampli\right \vert}{\left \vert \para[k]\right \vert}$. At this point, the integral in the propertime $T$ can be performed. 
In order to isolate the divergences in $D=4$, one can introduce the customary parameter $\mu$
to render $\ampli$ dimensionless and write
\begin{align}
 \begin{split}
\B
 &=-\frac{\mu }{2^{2 D-4} \pi ^{\frac{3 D-7}{2}}}\left(\frac{\ampli}{\mu}\right)^{5-D}  \Gamma\left(\frac{5-D}{2}\right) \int_0^1\frac{ ds}{\sqrt{s}} \int_0^{\infty} du \left[   \frac{\left(1-s\right)^{D/2-2}}{ (\aaa^{-2}s+ u+u^2)^{(5-D)/2}}\right.\\
 &\hspace{6.1cm}\left.-  \frac{\left(1-s\right)^{D/2-2}}{ (\frac{1}{2}+u)^{(5-D)}}+ \frac{\left(1-s\right)^{D/2-2}}{ ( \frac{1}{2}+u)^{(5-D)}}\right].  
 \end{split}
\end{align}
This leaves thus the singularity at $D=4$ unveiled:
\begin{align}
\begin{split}
  \B^{div}:&=-\frac{\mu}{2^{2 D-4} \pi ^{\frac{3 D-7}{2}}} \left(\frac{\ampli}{\mu}\right)^{5-D}\Gamma\left(\frac{5-D}{2}\right) \int_0^1\frac{ ds}{\sqrt{s}} \int_0^{\infty} du  \frac{\left(1-s\right)^{D/2-2}}{ ( u+\frac{1}{2})^{(5-D)}}\\
  &=-\frac{\mu}{2^{3 D-7} \pi ^{\frac{3 D}{2}-4}} \left(\frac{\ampli}{\mu}\right)^{5-D} \frac{\Gamma\left(\frac{5-D}{2}\right)\Gamma\left(\frac{D}{2}-1\right)}{\Gamma\left(\frac{D-1}{2}\right)} \frac{1}{(4-D)} \,  .
\end{split}
 \end{align}
The remaining contributions to $\B$ yield a finite contribution, as can be seen from a direct computation:
\begin{align}
 \B^{fin} :&=\frac{\ampli }{32\pi^2} \left( \left(2+\aaa\right) \log\left(1+\frac{2}{\aaa} \right)- 2-2\log 2 \right).
\end{align}

Lastly, we have the contribution from the product of two erfc functions:
\begin{align}
 \begin{split}
\C
 :&=\frac{\ampli^2 }{2^{2 D-5} \pi ^{\frac{3 D-9}{2}}}\int_0^{\infty} \frac{dT}{T^{\frac{D-3}{2}}}  \int_0^1 ds (1-s)   e^{ -k^2s \left(1-s\right)T  } \\
 &\hspace{3cm}\int^\infty_0\int^\infty_0 du_1du_2 K_0(u_1,0;T s) K_0(u_2,0;T (1-s)).  
 \end{split}
\end{align}
One can reverse the order in the integrals, and discover by direct integration that the result is convergent in $D=4$.
A closed expression for $C$ involves Lerch's transcendent function $\Phi$, to wit
\begin{align}
 \C&=\frac{\ampli a }{128\pi^2(1+a)} \Phi\left(\frac{1}{(1+a)^2},2,\frac{1}{2} \right).
\end{align}

Combining the above results, the form factor in $D=4$  for a massless field reads
\begin{align}\label{eq:ff_explicit}
\ff(\para[k],\ampli)&=\frac{\ampli}{16\pi^2}\frac{1}{D-4}+\ff_{L}(\para[k],\ampli)+\ff_{NL}(\para[k],\ampli),
\end{align}
where we have defined a finite local and a finite ``nonlocal'' 
contribution\footnote{The contribution that we are calling nonlocal, upon a series expansion, may actually contain some terms that are local.}:
\begin{align}
 \ff_{L}(\para[k],\ampli)&=\frac{\ampli }{32 \pi ^2} \left(\gamma -4+\log \left(\frac{\mu^2}{16 \pi ^3\ampli^2}\right)\right),\label{eq:ffl}\\
\ff_{NL}(\para[k],\ampli)&=\frac{\ampli}{32\pi^2}\left[-\frac{1}{\aaa}
+ \left(2+\aaa\right) \log\left(1+\frac{2}{\aaa} \right)+\frac{ \aaa }{4(1+\aaa)} \Phi\left(\frac{1}{(1+\aaa)^2},2,\frac{1}{2} \right)
\right]\,
,\label{eq:ffnl}
\end{align}
where  $\gamma$ is Euler's constant.

The divergent part of the form factor does not depend on $\para[k]$, 
and produces the following divergence of the effective action:
\begin{equation}\label{eq:divergence}
\Gamma^{(2)}_{div}=\frac{\ampli}{32\pi^2}\frac{1}{4-D}\int d\para\,  \eta^2(\para)\, .
\end{equation}
This result coincides with the one obtained in Ref. \cite{Bordag:2004rx}, where it is shown that the divergence
of the effective action for a potential of the form $v(\para)\delta(x^D-L)$
is proportional to the  integral over the surface of  $v^3(\para)$. In our case we have $v(\para)=\ampli +\eta(\para)$, 
and we are obtaining here the term that is quadratic  in $\eta(\para)$, with the correct coefficient. Consistently with this, we have also checked that the divergent part of the effective action
for an homogeneous plate reads
\begin{equation}\label{eq:divergence0}
\Gamma^{(0)}_{div}=\frac{\ampli^3}{96\pi^2}\frac{1}{4-D}\int d\para\,  \, .
\end{equation}

Because of these divergences, we should appeal to a renormalization process. 

\subsection{The renormalization}
\colblack{ As can be inferred from the divergence in Eq. \eqref{eq:divergence}, 
the $a_2$ coefficient\footnote{Or $a_4$: it depends on the convention used to number the coefficients, which may include semi-integers or only integers. In any case, we refer to the coefficient which in $D=4$ accompanies the zeroth power of the proper time.} of the HK in the Seeley-DeWitt expansion turned out to be nonvanishing. 
As usual, this implies that we should introduce a renormalization procedure. 
}

\colblack{
As explained in the introduction, a frequently used renormalization criterion 
consists in introducing counterterms containing classical fields. 
These counterterms are chosen in such a way that the divergences can be absorbed in the couplings of the theory
and the large mass limit for the renormalized energy gives a vanishing result \cite{Bordag:1998vs}. 
Given that we are considering a massless field, this option is not available in our case.
One may to avoid this by introducing a finite mass, renormalizing with the preceding criterion and afterwards taking the massless limit. 
However, as could be expected, one would usually find ficticious \colblack{logarithmic} divergences. 
}

\colblack{Let us be even more explicit. 
One can employ the expansions in Eq. \eqref{eq:kd_local_expansions} for small propertime $T$, 
and then replace in Eqs. \eqref{eq:kernel_gamma} and \eqref{eq:master} to obtain 
the large-mass expansion of the effective action.
Introducing the renormalization scale $\mu$ and considering an expansion around four dimensions, $D=4-\varepsilon$, we are lead to 
\begin{align}
 \label{eq:mass_expansion}
 \Gamma^{(2)}_{m=\infty}=\left[\frac{m}{64 \pi }+\frac{\ampli}{64 \pi ^2 }\left(\frac{2}{\varepsilon}+ \log \left(\frac{4\pi^3\mu^2}{m^2}\right)-\gamma\right)\right]\int d\para\,  \eta^2(\para)\, .
\end{align}
Then the naive subtraction and limit
\begin{align}
 \label{eq:large_mass_subtraction}
   \Gamma^{(2)}_{\text{ren},\,m=\infty}=\lim_{\varepsilon\rightarrow 0} \left(\Gamma^{(2)}- \Gamma^{(2)}_{m=\infty}\right)
\end{align}
would be well defined, as long as the field is not massless. 
If instead we try to take the massless limit of $\Gamma^{(2)}_{\text{ren},\,m=\infty}$, we then find the above-mentioned logarithmic divergence. 
The same behavior is observed, for example, in the case of a homogeneous delta potential using the results in \cite{Munoz-Castaneda:2013yga}.}

\colblack{
Notice also that, as explained in \cite{Bordag:1998vs},  
the coefficients accompanying the powers of $m$ in \eqref{eq:mass_expansion} can be related to the Seeley-deWitt coefficients of the massless operator's HK. 
This can be readily verified by comparing expression \eqref{eq:mass_expansion} with the coefficients listed in \cite{Bordag:1999ed}; 
as said before, the $\log(m^2)$ term corresponds to the $a_2$ HK coefficient, which is non-vanishing in our model.
}

\colblack{For our purposes, we will just introduce a counterterm for the divergent term}. Both the meaning of the divergent piece \eqref{eq:divergence} and its renormalization will depend on the specific initial field theory.
In particular the $\para[k]$-independent  term of the form factor will be defined only after fixing a renormalization prescription. If one sticks to the image of $\ampli$ being a coupling, then the divergence could be absorbed into a redefinition
of the mass of the field that describes the inhomogeneities. If instead one considers the theory previously described 
where $\ampli+\eta(\para)$ is the vacuum expectation value of another field, then the renormalization involves a cubic term. Calling $\mathcal{O}$ the composite operator whose coupling will absorb the divergence in Eq.\eqref{eq:divergence}, 
in the following we will just assume that the renormalization prescription is such that the couplings of all composite operators other than $\mathcal{O}$ can be taken to be their corresponding one-loop contributions\footnote{At least at a certain energy scale relevant to the problem.}. A further analysis of these aspects will be left to a future publication.

As a way to ascertain the validity of our results, 
notice that the form factor coincides, up to the ambiguous constant term, with the expression one obtains using the Gel'fand-Yaglom theorem. 
We include a sketch of this derivation in Appendix \ref{app:GY}, mimicking what is done in  \cite{Fosco:2019lmw} for the case of two interacting layers. 

This alternative calculation also \colblack{suggests} that the nonlocal part 
of the form factor does not depend on the regularization scheme\footnote{Similar resuts are obtained by regularizing with an UV momentum cutoff $\Lambda$ the propertime integrals,  $\int_0^\infty dT\to\int_{1/\Lambda^2}^\infty dT$. }. 
\colblack{ The uniqueness of general Casimir results, in the sense of regularization/renormalization independence, has been widely discussed in the bibliography \cite{Beneventano:1995fh, Kirsten:2001wz}.
Turning to our case, in addition to the preceding argument one can also give a HK explanation as follows.
}

\colblack{It has been shown in Ref. \cite{Bordag:1999ed} that the divergences arising from  the quantum fluctuations of a scalar particle in an arbitrary curved manifold of $D=4$ dimensions, subject to an interaction with curved delta plates, are given by a finite number of geometric invariants. This means that one will always be able to introduce just a finite number of counterterms to proceed with the renormalization. Whether this process will end up to be predictive or not will depend on the existence and interpretation of the finite terms.
}

\colblack{
For the case under consideration in this article, the possible divergent terms are four: $\int d\para \psi$, $\int d\para \psi^2$, $\int d\para \psi_{;aa}$ and $\int d\para \psi^3$, where $\psi(x)=\zeta+\eta(x)$ and $\psi_{;aa}$ means the covariant Laplacian on the plates. The first two of them vanish because we are using dimensional regularization and  the third one because it is just a boundary contribution. Regarding the fourth term, the $\eta$ contribution vanishes by assumption, the $\eta^2$ one is given in Eq. \eqref{eq:divergence}, while the remaining $\eta^3$ goes beyond the approximation used. 
}

\colblack{As an overall result, only the local term is subject to a renormalization and is therefore not a prediction of the theory. The nonlocal terms are indeed a prediction of the theory\footnote{As said before, strictly speaking this is valid if one does not impose additional finite renormalizations for these terms}, analogously to what happen with the $\log R$ term for the ground energy of a sphere \cite{Bordag:1998vs}. We will further discuss these aspects in Sec. \ref{sec:examples} and Sec. \ref{sec:DCE}, when we will investigate some examples in Euclidean and Minkowski space.
}

\subsection{Smooth and rapidly varying inhomogeneities}\label{sec:smooth_rapid} From Eqs.\eqref{eq:master} and \eqref{eq:form_factor} 
we see that, when the inhomogeneities are smooth, the effective action is dominated by  the small-$\para [k]$ limit of the form factor.
Conversely, short wavelengths are relevant for rapidly varying
inhomogeneities.  We will then study the behavior of the form factor in these two opposite limits,
noting that
the constant $\ampli$ provides the mass scale to compare with the
wavelengths of the inhomogeneities, choosing either $a\gg 1$ or $a\ll 1$. 

The expansion of the form factor for smoothly varying inhomogeneities is given by the expression
\begin{align}\label{eq:ffnl_rapid}
  \ff_{NL}(\para[k],\ampli)=\frac{3\ampli}{32\pi^2}\left[1-\frac{2}{27 \aaa^2}+\frac{8}{225 \aaa^4}-\frac{8}{135 \aaa^5}+\frac{176}{2205 \aaa^6}-\frac{32}{315 \aaa^7}+\mathcal{O}\left({\aaa}^{-8}\right)\right].
\end{align}
Recall that the leading term, i.e. the local contribution, would be involved in the renormalization process,
in which also the $F_{L}$ contribution of Eq. \eqref{eq:ffl} takes part. 

By looking into the first three terms of expansion \eqref{eq:ffnl_rapid}, 
one could have thought that it could have been made only of even powers of $a$, meaning that the expansion would have been in powers of the Laplacian 
in a so-called derivative expansion. Instead, the term $a^{-5}$ signals the presence of half-integer powers of $a$, which render the expression nonlocal.
They will be also crucial to the emergence of a nonvanishing vacuum decay probability
in the Minkowski case, which will be analyzed in Sec. \ref{sec:DCE}.
It is interesting to point out that a similar nonanalytic term,
proportional to ${\para[k]}^5$, appears in the self-energy of a Dirichlet mirror with small geometric deformations \cite{Fosco:2012mn}.

The nonlocality is also observed in the small-$a$ expansion of the form factor, which reads
\begin{align}\label{eq:ffnl_smooth}
  \ff_{NL}(\para[k],\ampli)&=\frac{\para[k]}{32\pi^2}\left[-1-2 \aaa \log \left(\frac{\aaa}{2}\right)+\aaa^2 \left(-\log \left(\frac{\aaa}{2}\right)+\frac{\pi ^2}{8}+1\right)+\mathcal{O}\left(\aaa^3\right)\right].
\end{align}
Indeed, there are integer and semi-integer positive powers of $k$ in such an expansion, as well as logarithmic terms that appear as soon as the inhomogeneity $\eta$ interacts with a constant distribution over the layer, i.e. whenever $\ampli$ is nonvanishing. Notice also that the leading term could have been guessed from a dimensional analysis. 
This is the reason why it appears also in other effects related 
to the Casimir energy, such as the roughness correction \cite{MaiaNeto:2012cu}.

\subsection{Gau{\ss}ian inhomogeneities}\label{sec:examples}
The preceding expressions can be readily used to analyze physical situations in which the inhomogeneities are time-independent. 
In that case, considering a four-dimensional Euclidean space, expression \eqref{eq:master} simply reduces to
\begin{align}\label{eq:time_independent}
\Gamma^{(2)}_{TI}:= -\frac{T_0}{2}\int \frac{d\paras[k]}{(2\pi)^{2}}  \, \vert \tilde\eta(\paras[k])\vert^2 \ff(\paras[k],\ampli),
\end{align}
where the relevant Fourier variables $\paras[k]$ are ``spatial'', 
and the length of the Euclidean time domain has been written as an overall factor $T_0$.

To be more explicit, consider now the particular case where the inhomogeneities have a Gau{\ss}ian isotropic form
\begin{align}
 \eta(\paras[x]):=\frac{\eta_0}{{2\pi\sigma^2}}e^{-\frac{(\paras[x]-\paras[x]_0)^2}{2\sigma^2}} ,
\end{align}
centered at $\paras[x]_0$ and with a width characterized by $\sigma$; 
$\eta_0$ is a parameter of length dimension that measures its amplitude. 
Aside from $\eta_0$, the overall normalization has been chosen such 
that its integration over the whole space $\paras[x]$ gives unity. 
Consequently its Fourier transform acquires the simple expression
\begin{align}
 \tilde\eta(\paras[k])=\eta_0e^{-\frac{{\paras[k]}^2\sigma^2}{2}-i\paras[k]\paras[ x_0]}.
\end{align}
Replacing this particular profile in Eq.\eqref{eq:time_independent} and discarding the divergent contribution 
as well as the local $\ff_{L}$ factor we get
\begin{align}\label{eq:tig}
\begin{split}
\frac{\Gamma^{(2)}_{TI}}{T_0}:%&= -\frac{\left|\eta_0\right|^2}{2({2\pi\sigma})^4} \int d\paras[k]  \, e^{-{\paras[k]}^2\sigma^2} \ff(\paras[k],\ampli)\\
&=-\frac{\left|\eta_0\right|^2}{2({2\pi})^2}\int d\paras[k]  \, e^{-{\paras[k]}^2\sigma^2} \ff_{NL}(\paras[k],\ampli).
\end{split}
\end{align}
This choice corresponds to a particular renormalization condition, on which our results will depend; 
we will come back to this issue in the end of this section.

\begin{figure}[h]
\begin{center}
 \begin{minipage}{0.48\textwidth}
 \includegraphics[width=1.0\textwidth,height=0.8\textwidth]{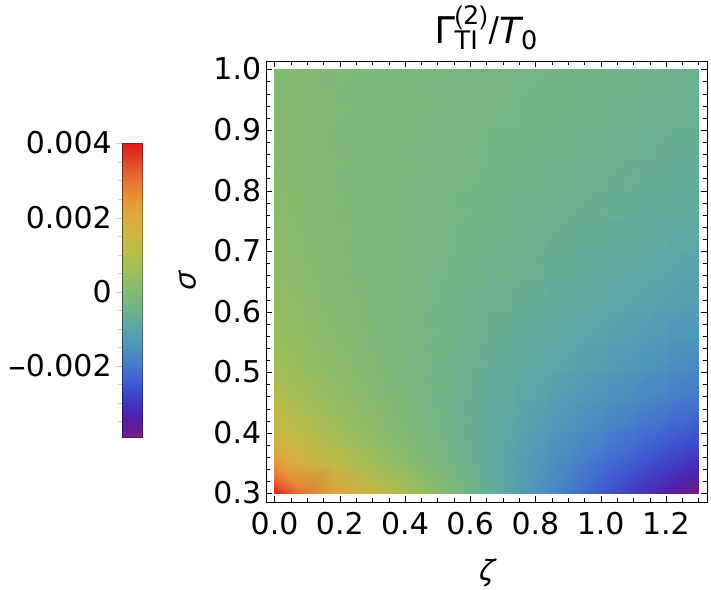}
 \end{minipage}
 \hspace{0.02\textwidth}
 \begin{minipage}{0.48\textwidth}
 \vspace{0.4cm}\includegraphics[width=0.95\textwidth]{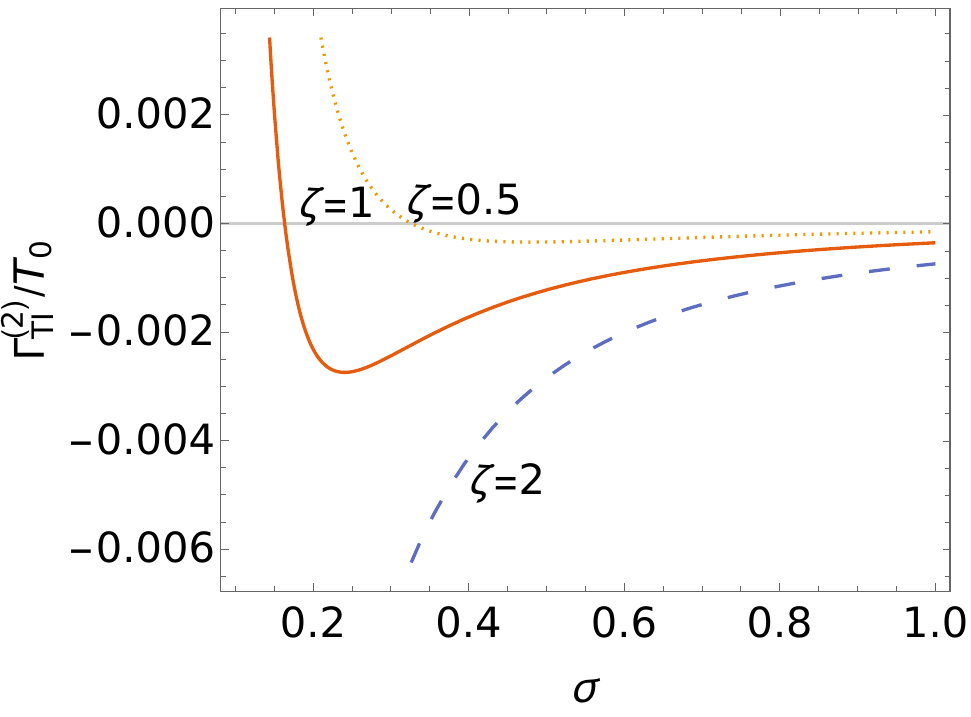}
 \end{minipage}
 \caption{ Contribution $\Gamma^{(2)}_{TI}$ to the effective action per unit time for Gau{\ss}ian inhomogeneities (in colors in the online version): the left panel is a density plot as a function of $\ampli$ and $\sigma$, while the right panel corresponds to a plot as a function of $\sigma$ for several values of $\ampli$.
 All the dimensional quantities are measured in arbitrary units, and we have chosen $\eta_0=1$.}
 \label{fig:tig}
 \end{center}
\end{figure}

Even if a closed expression for the integral in Eq.\eqref{eq:tig} is not available, a numerical integration can be readily performed. 
On the left panel of Fig. \ref{fig:tig} we show a density plot of the contribution $\Gamma^{(2)}_{TI}$ 
per unit time as a function of $\sigma$ and $\ampli$, 
while in the right panel we plot it as a function of $\sigma$ for several values of $\ampli$, 
setting in both cases the amplitude $\eta_0$ to one.

First of all, notice that $\Gamma^{(2)}_{TI}$  increases as $\sigma$ tends to zero,
namely when the first term in the RHS of Eq.\eqref{eq:ffnl_smooth} prevails.
This is to some extent hidden in the right panel of Fig. \ref{fig:tig} for large values of $\ampli$, 
because the negative minimum also gets shifted towards smaller values of $\sigma$.
Since in the small $\sigma$ limit the Gau{\ss}ian becomes a delta function, 
the divergence is to be expected for two reasons:
neither the interaction with a point delta potential in $D>2$ is well defined, 
nor the product $\eta^2$ which would involve two deltas. 

\begin{figure}[h]
\begin{center}
 \begin{minipage}{0.48\textwidth}
 \includegraphics[width=1.0\textwidth,height=0.8\textwidth]{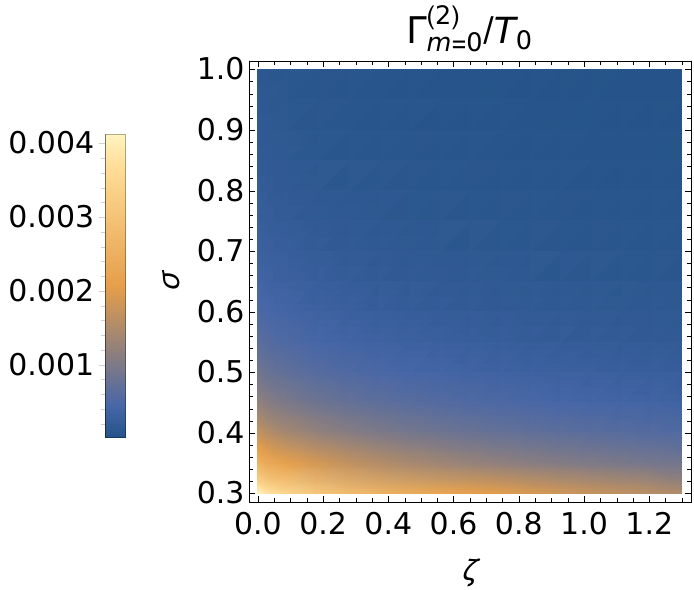}
 \end{minipage}
 \hspace{0.02\textwidth}
 \begin{minipage}{0.48\textwidth}
 \vspace{0.4cm}\includegraphics[width=0.95\textwidth]{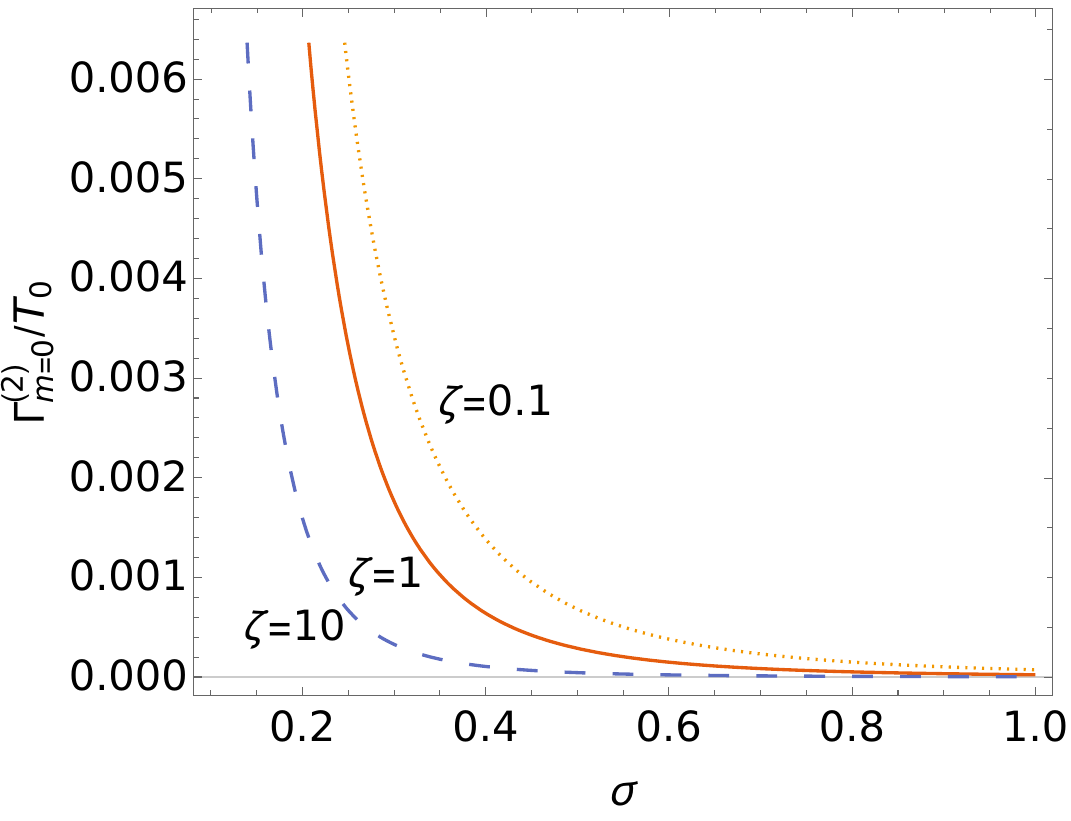}
 \end{minipage}
 \caption{ Nonlocal contribution to the effective action (in the ``null-mass'' renormalization) per unit time ($\frac{\Gamma^{(2)}_{m=0}}{T_0}$) for Gau{\ss}ian inhomogeneities (in colors in the online version): the left panel is a density plot as a function of $\ampli$ and $\sigma$, while the right panel corresponds to a plot as a function of $\sigma$ for several values of $\ampli$.
 All the dimensional quantities are measured in arbitrary units, and we have chosen $\eta_0=1$.}
 \label{fig:tigNL}
 \end{center}
\end{figure}

Secondly, for fixed $\ampli$ and large $\sigma$, the value of $\Gamma^{(2)}_{TI}$ tends to zero. 
This is merely related to the normalization chosen for the Gau{\ss}ian.
Additionally, we observe that for fixed $\sigma$ the contribution changes sign as $\ampli$ increases,
tending to a linear behavior. This agrees with the  explicit expansion of $\Gamma^{(2)}_{TI}$ for small and large $\sigma\ampli$:
\begin{align}
\frac{\Gamma^{(2)}_{TI}}{T_0}=
\begin{cases}
\displaystyle \frac{1}{512 \pi ^3 \sigma^3} \Big(\sqrt{\pi }+4 \log (\sigma\ampli)\sigma \ampli +2 (\gamma -\log (4)) \sigma \ampli +\cdots\Big),\quad &\sigma\ampli\ll1\\[0.3cm]
 \displaystyle \frac{1}{\sigma^3}\left(-\frac{3 \sigma  \ampli}{256 \pi ^3}+\frac{1}{1152 \pi ^3 \sigma  \ampli}-\frac{1}{1200 \pi ^3 \sigma ^3 \ampli^3}+\cdots\right),\quad &\sigma\ampli\gg1
\end{cases}.
\end{align}

Nevertheless, this fact heavily depends on the chosen renormalization condition.
As a way to clarify this assertion, consider the subtraction 
\begin{align}\label{eq:tig2}
\begin{split}
\frac{\Gamma^{(2)}_{m=0}}{T_0}:%&= -\frac{\left|\eta_0\right|^2}{2({2\pi\sigma})^4} \int d\paras[k]  \, e^{-{\paras[k]}^2\sigma^2} \ff(\paras[k],\ampli)\\
&=-\frac{\left|\eta_0\right|^2}{2({2\pi})^2}\int d\paras[k]  \, e^{-{\paras[k]}^2\sigma^2} \left(\ff_{NL}(\paras[k],\ampli)-\frac{3\ampli}{32\pi^2}\right),
\end{split}
\end{align}
which corresponds to a renormalization that will be called ``null-mass'' condition.
Analogous to the previous case, we include in the left panel of Fig. \ref{fig:tigNL} a density plot of $\Gamma^{(2)}_{m=0}$ per unit time
as a function of $\ampli$ and $\sigma$, while the right panel corresponds to a plot as a function of $\sigma$ for different values of $\ampli$.
From them it can be seen that, contrary to the precedent situation, the contribution to the effective action is strictly positive and shows no local minima.
As emphasized before, this does not imply that the effective action has no physical meaning; 
it just points out the fact that one should formulate a theory describing the degrees of freedom encoded in $\eta$ and 
state a consistent renormalization prescription.

\section{Wick rotation and dynamical Casimir effect}\label{sec:DCE}
Time-dependent inhomogeneities excite the quantum vacuum,  with the subsequent particle creation. A quantity that measures the 
dissipative effects of the external time-dependent conditions on the quantum field is the vacuum persistence amplitude,  that can be computed from the effective action $\Gamma_M$ in Minkowski space
as follows
\begin{equation}
\langle 0_{out}\vert0_{in}\rangle=e^{i\Gamma_M}\, .
\end{equation}
The probability of pair creation $P$  is determined by the 
imaginary part of the effective action
\begin{equation}
1-P=\vert \langle 0_{out}\vert0_{in}\rangle\vert^2=e^{-2\operatorname{Im} \Gamma_M}\, .
\end{equation}
When the effective action is computed perturbatively we have, to lowest order,  $P\simeq 2\operatorname{Im}\Gamma_M$.

Coming back to to our model, one of the main advantages of the master formula \eqref{eq:master} 
is that it admits a Wick rotation from Euclidean to Minkowski space.
Indeed, one can show that a Wick rotation $x^0\rightarrow ix^0$ encounters no singularities in the complex plane; 
the resulting effective action $\Gamma_M$ in Minkowski space is
\begin{align}\label{GammaM}
 \Gamma_M^{(2)} &=\frac{1}{2}\int \frac{d\paras[k]dk^0}{(2\pi)^3}\vert \eta_M(k^0,\paras[k])\vert^2 \ff(\para[k]_M,\ampli), 
\end{align}
where the Fourier transform shall be defined as
\begin{align}
 \eta_M(\para[x])=\int \frac{d\paras[k] dk^0}{(2\pi)^3} e^{-ik^0x^0+i\paras[k] \paras[x]} \eta_M(k^0,\paras[k]),
\end{align}
and the expressions involving the norm of $\para[k]_E$ in Euclidean space should be understood by 
introducing a negative imaginary part inside the square root, i.e. $\vert \para[k]_E \vert\rightarrow \vert\para[k]_M\vert=\sqrt{{\paras[k]}^2-k_0^2-i0}$.

Taking into account that the form factor  $\ff$ is a real function
of the Euclidean momentum, the imaginary part of the Minkowskian effective action reads
\begin{align}\label{eq:master_vacuum}
\begin{split}
\operatorname{Im} \Gamma_M^{(2)}&=\frac{1}{2}\int \frac{d\paras[k] dk^0}{(2\pi)^3}\vert\eta_M(k^0,\paras[k])\vert^2\theta\left( (k^0)^2-{\paras[k]}^2\right)\\
&\hspace{3.5cm}\times\operatorname{Im} \left (\ff\left (-i\sqrt{(k^0)^2-{\paras[k]}^2},\ampli\right)\right)
\end{split}
\end{align}
in terms of the Heaviside function $\theta$.
Expression \eqref{eq:master_vacuum} is the analog in our problem of well-known formulas for the vacuum persistence amplitude 
(consider for example quantum fields in the presence of classical electromagnetic fields \cite{Itzykson}, 
or quantum fields on curved spacetimes \cite{Frieman:1985fr}).
Notice that this expression is independent of the renormalization prescription, which only involves a real and local term. Additionally, for massive quantum fields of mass $m^2$ the imaginary part will contain a factor $ \theta\left( (k^0)^2-{\paras[k]}^2-4 m^2\right)$, 
which is nothing but the threshold for pair production. 

\subsection{Gau{\ss}ian inhomogeneities}
To show the plasticity of these formulas in Min\-kows\-ki space, consider once more 
a Gau{\ss}ian toy model,
\begin{align}
\eta_G(\para[x])= \eta_0\frac{e^{\frac{-(\paras[x]-\paras[y])^2}{2\sigma_s^2}}}{2\pi\sigma_s^2}\frac{e^{\frac{-(x^0-y^0)^2}{2\sigma_t^2}}}{\sqrt{2\pi\sigma_t^2}},
\end{align}
so that $\eta_0$ is a parameter of dimension length square in $D=4$ and, expliciting the Minkowskian metric, 
its Fourier transform reads
\begin{align}\label{eq:eta_gm}
 \eta_G(\para[k])=\eta_0 e^{-\frac{{\paras[k]}^2\sigma_s^2}{2}-\frac{(k^0)^2\sigma_t^2}{2}-i\paras[k] \paras[y]+ik^0 y^0}.
\end{align}

For a moment, consider the term which accounts only for the interaction among the inhomogeneities $\eta$, 
namely set $\ampli=0$. A straightforward replacement in Eq. \eqref{GammaM} gives the following formula for the effective action 
\begin{align}\label{eq:gamma_mg}
 \begin{split}
\left.\Gamma_{M,G}^{(2)}\right\vert_{\ampli=0}
 %&=- \frac{\vert\eta_0\vert^2 }{16 (2 \pi)^{5}}  \int {d\paras[k] dk^0} e^{-{\paras[k]}^2\sigma_s^2-(k^0)^2\sigma^2_t} \sqrt{{\paras[k]}^2-(k^0)^2-i\epsilon}\\
 &=-\frac{\vert\eta_0\vert^2 }{2^6 (2 \pi)^{4}} \left[\frac{\pi }{ \sigma_s^3 \sqrt{\sigma_s^2+\sigma_t^2}}  -\frac{2i}{ \sigma_s^3 \sigma_t^2} \left( \sigma_s-\frac{\sigma_t\sinh ^{-1}\left(\frac{\sigma_s}{\sigma_t}\right)}{\sqrt{\frac{\sigma_s^2}{\sigma_t^2}+1}}\right)\right],
 \end{split}
 \end{align}
from which the following expansions for the vacuum persistence amplitude can be immediately obtained:
 \begin{align}\label{eq:vacuum_mg}
\left.\operatorname{Im}\Gamma_{M,G}^{(2)}\right\vert_{\ampli=0}
 &\sim\frac{\vert\eta_0\vert^2 }{2^4 (2 \pi)^{4}}
 \begin{cases}
  \frac{1}{2\sigma_s^2\sigma_t^2}\left[1+ \log\left(\frac{\sigma_t}{2\sigma_s}\right) \left(\frac{\sigma_t}{\sigma_s}\right)^2+\cdots \right], &\sigma_t\ll \sigma_s\\[0.2cm]
  \frac{1}{3\sigma_t^4}\left[1-\frac{4}{5} \left(\frac{\sigma_s}{\sigma_t}\right)^2+\cdots\right], &\sigma_s\ll \sigma_t\\[0.3cm]
 \frac{2-\sqrt{2}\,\text{arcsinh}(1)}{4} \frac{1}{\sigma^4}\approx\frac{0.0941937}{\sigma^{4}}, &\sigma_t=\sigma_s=\sigma
 \end{cases}.
\end{align}
Notice first of all that $\left.\operatorname{Im}\Gamma_{M,G}^{(2)}\right\vert_{\ampli=0}$ is always positive, as can be proved from expression \eqref{eq:gamma_mg}. 
Secondly, Eq.\eqref{eq:vacuum_mg} means that for inhomogeneities highly localized in time we obtain a divergent contribution for the imaginary part, 
while the real part real part remains finite, as could be expected for a sudden perturbation. 
The opposite situation is obtained in the case of spatially-localized inhomogeneties, 
where the real part diverges, remaining the imaginary part finite.
\begin{figure}[h]
\begin{center}
\begin{minipage}{0.42\textwidth}
 \vspace{0.1cm}\includegraphics[width=1.0\textwidth]{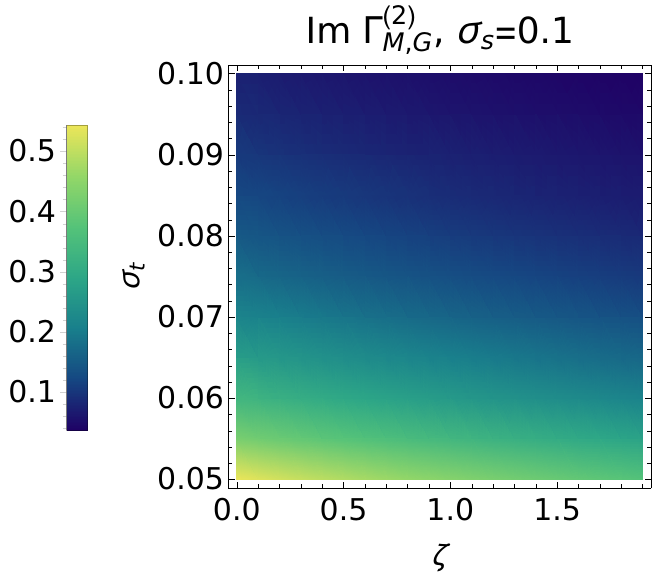}
 \end{minipage}
 \hspace{0.02\textwidth}
  \begin{minipage}{0.42\textwidth}
 \vspace{0.1cm}\includegraphics[width=1.0\textwidth]{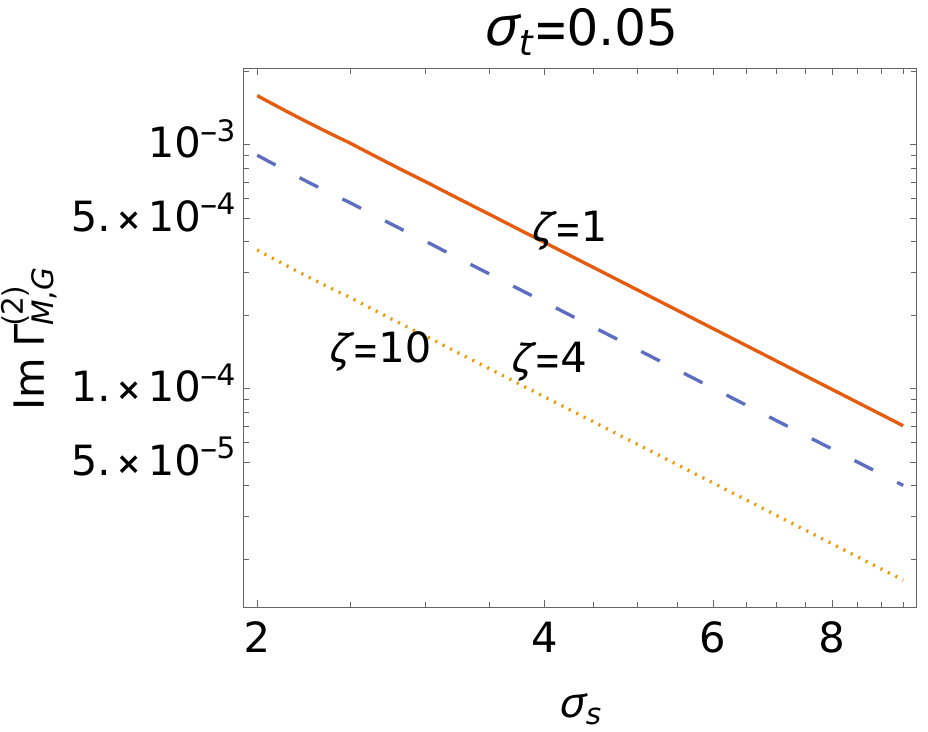}
 \end{minipage}

\caption{ Imaginary part of $\Gamma_{M,G}^{(2)}$ for Gau{\ss}ian inhomogeneities (in colors in the online version).
The left panel corresponds to a density plot as a function of $\ampli$ and $\sigma_t$ for $\sigma_s=0.1$,
while the right panel is a double-log plot as a function of $\sigma_s$ for fixed $\ampli=1.$ and $\sigma_t=0.05$. 
All the dimensional quantities are measured in arbitrary units, and $\eta_0=1$ is chosen.}
 \label{fig:vacuum_g}
 \end{center}
\end{figure}

Turning our attention to the complete effective action \eqref{GammaM} with Gau{\ss}ian inhomogeneities,
the numerical computations can be easily handled.
In particular, in Fig. \ref{fig:vacuum_g} we show the imaginary part of
 $\Gamma_{M,G}^{(2)}$: 
the left panel corresponds to a density plot as a function of $\ampli$ and $\sigma_t$ (for fixed $\sigma_s$),
while the right panel is a double-log plot as a function of $\sigma_s$ (for fixed $\ampli$ and $\sigma_t$).

Once more, we observe that $\operatorname{Im}\Gamma_{M,G}^{(2)}$ is positive, as it should be according to its interpretation.
Next, it is important to notice that $\Gamma_{M,G}^{(2)}$ 
is a decreasing function with respect to all the involved variables, 
with asymptotic power law behaviors.
As an example, in the right panel of Fig. \ref{fig:vacuum_g} we depict its $\sigma_s^{-2}$ 
asymptotic behavior for large $\sigma_s$.
In spite of that, the nature of these decreases is decidedly different:
while the decrease with $\ampli$ for a regular $\eta$ could be ultimately related 
to the $a^{-5}$ factor in the expansion \eqref{eq:ffnl_rapid} (see also the next section), 
the behavior with $\sigma_{s,t}$ heavily depends on the chosen inhomogeneities. 

Last, $\operatorname{Im}\Gamma_{M,G}^{(2)}$ diverges for small $\sigma_t$ but converges for small $\sigma_s$, that is, the delta limit of the space distribution is well defined for this quantity.

\subsection{Harmonic inhomogeneities}
Let us now focus on pertubations $\eta$ which are homogeneous in space
and have a harmonic time dependence. 
In particular we will consider a generalization, to an arbitrary dimension $D$,
of the model proposed in \cite{Silva:2011fq} to simulate the dynamical Casimir effect.
We introduce thus
\begin{align}
 \eta_H(t)&=\eta_0 \cos\left(\omega_0 t\right) e^{-\frac{\vert t\vert }{T}},
\end{align}
a harmonic inhomogeneity with frequency $\omega_0$ whose amplitude 
is modulated by an exponential as a way to regularize the expressions;
$\eta_0$ has dimensions of mass and
in the following we will take the limit of large $T$.
 Under this assumption its Fourier transform simplifies, yielding
\begin{align}
\eta_H(k^0)\eta_H(-k^0)&=\frac{\pi}{2}\vert\eta_0\vert^2 T \left[\delta(k^0-\omega_0)+\delta(k^0+\omega_0)\right],\quad  \omega_0T\gg1,
\end{align}
and rendering the evaluation of the effective action in Eq.\eqref{GammaM} immediate.
Given that the effective action becomes proportional to $A$, the spatial area of the surfaces where the inhomogeneities live,
as well as to the characteristic time $T$, 
 we find it more appropriate to consider the expression per unit time and area, 
\begin{align}\label{eq:gamma_h}
 \frac{\Gamma^{(2)}_{M,H}}{A\, T}
 &= \frac{\vert\eta_0\vert^2}{4}  \ff(-i\omega_0,\ampli).
 \end{align}
As physically guessed, we can see that the behavior will be fixed by the relative size of the two scales of the system, $\omega_0$ and $\ampli$.
Considering the imaginary part of \eqref{eq:gamma_h}, 
relevant for the computation of the decay rate of the vacuum per unit time and area of the delta sheets,
we obtain the following expansions:
\begin{align}\label{eq:expansion_h}
\frac{\operatorname{Im} \left(\Gamma^{(2)}_{M,H}\right)}{A\,T}= \frac{\vert\eta_0\vert^2 \omega_0}{16\pi^2}
 \begin{cases}
         \frac{1}{8}\left[1-\frac{\pi  \ampli}{\omega _0}+\left( \log \left(\frac{2\omega _0}{\ampli}\right)+\frac{\pi}{8} ^2+1\right)\frac{\ampli^2}{ \omega _0^2}
         %-\frac{\pi  }{4}\frac{\ampli^3}{ \omega _0^3} 
         +\cdots\right], & \ampli \ll\omega_0\\[0.3cm]
          \frac{1}{45}\frac{\omega _0^4}{ \ampli^4}\left[1-\frac{12 \omega _0^2}{7 \ampli^2}+\frac{96 \omega _0^4}{35 \ampli^4}+\cdots\right], &\ampli\gg\omega_0
         \end{cases}.
\end{align}

\begin{figure}[h]
\begin{center}
 \begin{minipage}{0.42\textwidth}
 \vspace{0.1cm}\includegraphics[width=1.0\textwidth]{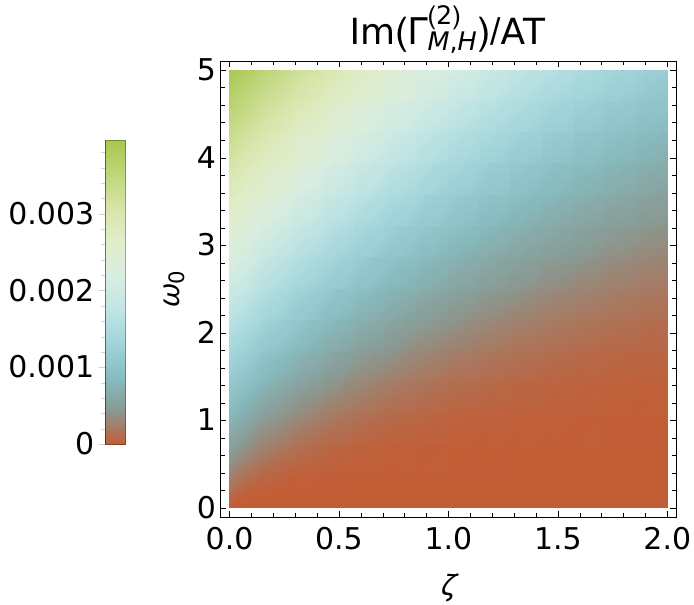} 
 \end{minipage}
\hspace{0.02\textwidth}
\begin{minipage}{0.42\textwidth}
 \vspace{0.1cm}\includegraphics[width=1.0\textwidth]{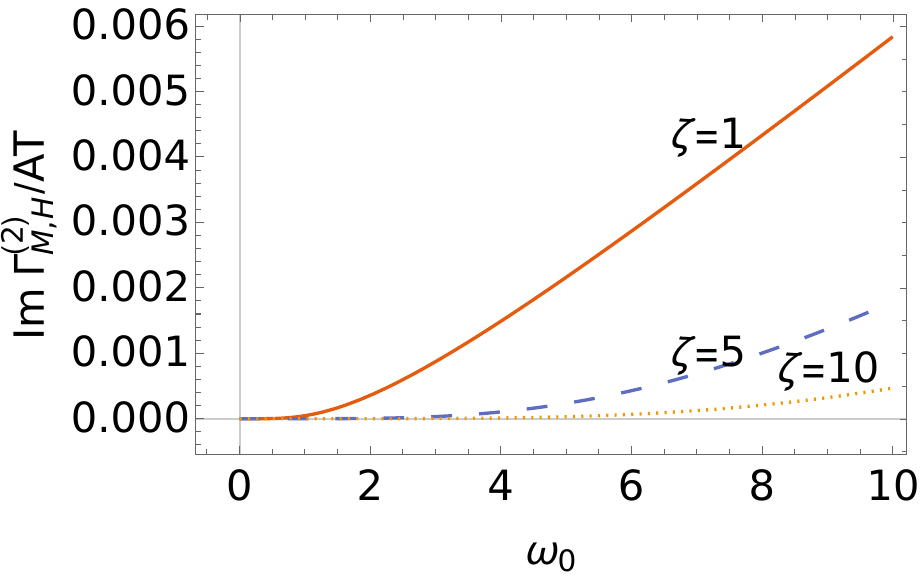} 
 \end{minipage}
 
\caption{ Plot of ${\operatorname{Im} \left(\Gamma^{(2)}_{M,H}\right)}$ per unit time and area
for harmonic inhomogeneities (in colors in the online version).
The left panel corresponds to a density plot as a function of $\ampli$ and $\omega_0$,
while the right panel is a plot as a function of $\omega_0$ for several values of $\ampli$. 
All the dimensional quantities are measured in arbitrary units and $\eta_0=1$ is chosen.}
 \label{fig:vacuum_h}
 \end{center}
\end{figure}

The qualitative behaviors of the expansions in Eq.\eqref{eq:expansion_h} could have certainly been guessed by looking at
the formulas given in Sec. \ref{sec:smooth_rapid} for the form factor:
for large $\omega_0$ the main term is the first in Eq.\eqref{eq:ffnl_rapid}, 
given only by the geometry of the inhomogeneities, 
while for small $\omega_0$ the first term corresponds to the first 
contribution in Eq.\eqref{eq:ffnl_smooth} that prevented a derivative expansion 
(the $a^{-5}$ term). 
Regarding the latter, 
it is suggestive to notice its analogy with the 
imaginary part of the effective action for a single perfect mirror in the usual dynamical Casimir effect configuration \cite{Fosco:2007nz},
which is also proportional to 
$\omega_0^5$ (notice that to perform a correct comparison, one should take the appropriate scaling $\eta_0\sim \ampli^2$).

To obtain exact results regarding the effective action, one can rely on numerical computations. 
Those shown in Fig. \ref{fig:vacuum_h} agree with the expansions in Eq.\eqref{eq:expansion_h}. 
In effect, the density plot in its left panel, 
which corresponds to $\operatorname{Im} \Gamma^{(2)}_{M,H}$ per unit time and area as a function of $\zeta$ and $\omega_0$, 
shows a decreasing (increasing) function of $\ampli$ ($\omega_0$) respectively. 
In addition, the right panel in Fig. \ref{fig:vacuum_h}, displays its plot
as a function of $\omega_0$, for several values of $\ampli$. 
The power law and the linear behavior, corresponding to small and large values of $\omega_0$, 
are clearly depicted.

\section{Conclusions}\label{sec:conclus}
In this paper we studied the effective action for a quantum scalar field in the presence of a thin and inhomogeneous layer. After presenting a novel derivation of the exact HK for the homogeneous case, we computed the effective action using a perturbative approach in the inhomogeneities. 
The general expression for a massive scalar field in $D$ dimensions was obtained, viz. Eq.\eqref{eq:master}, showing that is not well defined 
for $D\geq 3$. 

We analyzed in detail the case of a massless field in $D=4$ using dimensional regularization, for which we showed that the divergent term is local in the inhomogeneities, while the remaining finite part of the effective action is nonlocal both for smooth and rapidly varying inhomogeneities. We performed some cross-checks of our calculations, comparing the divergent part with the general results
obtained in
Ref.\cite{Bordag:2004rx}, and computing the finite nonlocal part
using a different approach based on the Gel'fand-Yaglom theorem \cite{Fosco:2019lmw}. 
\colblack{Judging by our results, the nonlocal part seems to be regularization independent. 
In doing so we also discussed some technical aspects regarding the renormalization for massless fields, 
which basically implie that the local term is not a prediction of the model.}

For time-independent inhomogeneities, the effective action gives the vacuum self-energy of the mirror. We analyzed the dependence of the effective action with the parameters of Gau\ss ian inhomogeneities.
We then considered spacetime-dependent inhomogeneities. In this case, we applied our results to compute the vacuum persistence amplitude, which is determined by the imaginary part of the Lorentzian effective action. After a Wick rotation of the Euclidean effective action, we obtained 
a general formula for the vacuum persistence amplitude, which was applied to several examples of interest in the context of the dynamical Casimir effect. The imaginary part of the Wick-rotated
effective action is always finite and positive definite, as expected.
Since the calculation of the vacuum persistence amplitude for spatially inhomogeneous mirrors with time-dependent properties can be employed 
to model the particle creation produced by mirrors of finite size, we expect that our general formulas could be of practical application.

More generally, the HK approach used in this paper is a suitable tool to compute local quantities, 
as the expectation value of the stress tensor. This is a compelling topic for future research, since we expect divergences in the renormalized stress tensor when evaluated near
the thin layer. These divergences should depend on the local inhomogeneities, and would be similar to those appearing for nonplanar geometries
with perfect boundary conditions, that depend on the 
local curvature \cite{Deutsch:1978sc}. In this scenario, the consideration of combined
effects produced by  geometry and inhomogeneities 
deserves further attention.

Finally, in order to go beyond the employed $\eta^2$ approximation, nonperturbative numerical computations following the lines of \cite{Franchino-Vinas:2019udt} could be performed. 
Some of these ideas are currently being pursued.

\section*{Acknowledgments}
S.A.F. is grateful to G. Gori and the Institut für Theoretische Physik, Heidelberg, for their kind hospitality.
SAF acknowledges support by UNLP, under project grant X909 and ``Subsidio a Jóvenes Investigadores 2019''. F.D.M. was supported by ANPCyT, CONICET, and UNCuyo.

\appendix
\section{Equivalence with the expression for $\Kd$ obtained in \cite{Gaveau_1986} }\label{app:Bauch}
In order to prove that formula \eqref{eq:HK_1d} is indeed equivalent to the one
obtained in \cite{Gaveau_1986} (up to a rescaling $\ampli=2a$, $T=2t$, where $a$ and $t$ correspond to the notation in \cite{Gaveau_1986}), we recast the $\ampli$-dependent part of $\Kd$ as
\begin{align}\label{eq:convolution1}
  \Delta K(T):&=\Kd(x,y;T)- K_0(x,y;T)\\
   &=  \bigg(K_0(x,\posi; \cdot) * \itkernel(\cdot) * K_0(\posi, y; \cdot) \bigg)(T),
\end{align}
where the operator $*$ means the ``Laplacian'' convolution,
\begin{align}
 (f*g)(t) =\int^t_0 d\tau f(\tau) g(t-\tau).
\end{align}
Then one can consider $\Delta \tilde K(s)$, the Laplace transform of $\Delta K(T)$; according to
the convolution theorem of the Laplace transform, it is simply given by the Laplace transform of the functions involved in the convolution
\eqref{eq:convolution1},
\begin{align}
 \begin{split}
\Delta \tilde K(s)&= \tilde K_0(x,\posi; s)  \tilde\itkernelzero_2(s) \tilde K_0(\posi, y; s)\\
 &=\ampli e^{-\sqrt{s}(|x-L|+|y-L|)} \frac{1}{2\sqrt{s}(\ampli+2\sqrt{s})}.
 \end{split}
\end{align}

Consider now a $\ampli$ such that $\text{Re}(\ampli)>0$. Antitransforming $\Delta \tilde K(s)$ we obtain 
\begin{align}
 \Delta K(T)=\frac{\ampli}{2\pi i} \int_{\gamma-i\infty}^{\gamma+i\infty} ds \frac{e^{sT-\sqrt{s}(|x-L|+|y-L|)}}{2\sqrt{s}( \ampli+2\sqrt{s})},
\end{align}
where $\gamma$ is such that all the singularities of the integrand lie on $\text{Re}(s)<\gamma$.
We can then deform the contour in the complex plane to encircle the half-line $(-\infty,\,\ampli^2)$, where the square root has a cut. 
After a  change of variables $w^2=s$, we use Schwinger's trick
\begin{align}
 \frac{1}{\ampli +iw}=\int^\infty_0 ds_1\, e^{-s_1(\ampli+iw)},
\end{align}
and perform the integral in $w$. Thus, considering the analytic continuation in $\ampli$, we obtain the promised result
\begin{align}
 \Delta K(T)=\frac{1}{2}\int^\infty_0 du\, e^{-\frac{u \ampli}{2}} K_0(|x-L|+|y-L|+u,0;T),
\end{align}
or equivalently
\begin{align}
 \Kd(x,y;T)
 &= K_0(x,y;T)- \frac{\ampli }{2}\int^\infty_0 du e^{-\frac{u \ampli}{2}} K_0(|x-L|+|y-L|+u,0;T).
\end{align}

\section{A direct proof of the convolution property \eqref{eq:HK_convolution}}\label{app:convolution}

One of the benefits of Eq. \eqref{eq:HK_1d} is that it provides a direct proof of the  convolution property \eqref{eq:HK_convolution}.
Since we  haven't found such a proof in the literature\footnote{By direct we mean using the explicit expression for the propagator.},
we include it in this appendix. 
As a first step, let us define a reduced iterated kernel as
\begin{align}\label{eq:identities_kernel}
 \itkernel(z_2-z_1) :&= \itkernelc(z_1-z_2)-\delta(z_1-z_2).
\end{align}
From this definition and considering formula \eqref{eq:iterated_kernel}, 
it is immediate to prove the following important identities, which we will call iteration properties:
\begin{align}\label{eq:iterated_props}
 \itkernel(z_3-z_1)&=
 \begin{cases}
\displaystyle \frac{\ampli}{\sqrt{4\pi}} \int_{z_1}^{T_2}  dz_2 \frac{1}{\sqrt{z_2-z_1}} \itkernelc(z_3-z_2),\quad \text{if } z_1<  z_3< T_2 \\[0.4cm]
\displaystyle \frac{\ampli}{\sqrt{4\pi}} \int_{T_1}^{z_3}  dz_2 \frac{1}{\sqrt{z_3-z_2}} \itkernelc(z_2-z_1),\quad \text{if }  T_1<z_1< z_3\\
\end{cases}.
\end{align}

Now notice that, after employing \eqref{eq:HK_1d}, the product in the RHS of \eqref{eq:HK_convolution} involves four terms. 
The first of them, upon using the convolution property of the free propagator,
becomes
\begin{align}
 \int dy \,K_0(x,y,T) K_0(y,z,S) =K_0(x,z,T+S).
\end{align}
The remaining three terms will instead combine in such a way that they reproduce the desired contribution, namely
\begin{align}\label{eq:concatenation_summed}
  \ampli \int_{0}^{T+S} dz_1 dz_2  K_0(x,L, z_1)\,\itkernelc( z_2-z_1) K_0(L,z, T+S-z_2).
\end{align}
In order to prove so, one term of the product in the RHS of Eq. \eqref{eq:HK_convolution} --still considering the replacement given by Eq.\eqref{eq:HK_1d}--
will just be simplified using the convolution of the free propagators:
 \begin{align}\label{eq:concatenation_term1}
  \ampli \int_{0}^{T} dz_1 dz_2  K_0(x,L, z_1)\,\itkernelc( z_2-z_1) K_0(L,z, T+S-z_2).
  \end{align}
In another one, we will also perform a translation $z_3,z_4\rightarrow z_3+T,z_4+T$
after convoluting the involved free propagators:
  \begin{align}
  \begin{split}\label{eq:concatenation_term2}
\ampli \int_{0}^{S}& dz_3 dz_4 \, K_0(x,L, z_3+T)\,\itkernelc( z_4-z_3) K_0(L,z, S-z_4)\\
  &=\ampli \int_{T}^{T+S} dz_3 dz_4  K_0(x,L, z_3)\,\itkernelc( z_4- z_3 ) K_0(L,z, T+S-z_4).   
  \end{split}
 \end{align}
 The remaining contribution requires more steps to be brought to the required form. 
 In a summarized way, we have
\begin{align}\label{eq:last_term_convolution}
\begin{split}
\ampli^2 &\int_{0}^{T} dz_1 dz_2 \int_{T}^{T+S}dz_3 dz_4  K_0(x,L, z_1)\,\itkernelc( z_2-z_1) K_0(L,L, z_3-z_2)\,\\
 &\hspace{6cm}\times\itkernelc( z_4-z_3) K_0(L,z, T+S-z_4)  \\
 &=\frac{\ampli^2}{\sqrt{4\pi}} \int_{0}^{T} dz_1 \int_{T}^{T+S} dz_4  \int_{0}^{T} dz_2 \left( \int_{z_2}^{T+S}dz_3 - \int_{z_2}^{T}dz_3 \right) K_0(x,L, z_1) \,\\
 &\hspace{2.5cm}\times\itkernelc( z_2-z_1 )\frac{1}{\sqrt{z_3-z_2}}\itkernelc( z_4-z_3) K_0(L,z, T+S-z_4)  \\
 &=\ampli\int_{0}^{T} dz_1 \int_{T}^{T+S} dz_4  K_0(x,L, z_1) K_0(L,z, T+S-z_4) \\
 &\hspace{0.5cm}\times\left( \int_{0}^{T} dz_2 \itkernelc( z_2-z_1) \itkernel(  z_4-z_2)   - \int_{0}^{T}dz_3 \,\itkernel( z_3-z_1) \itkernelc(  z_4-z_3)  \right) ,
 \end{split}
 \end{align}
 where in the second line we have made explicit the HK for coincident forms,
 while in the third line we have employed the iteration properties \eqref{eq:iterated_props}.
 Finally, considering Eq.\eqref{eq:identities_kernel} we have 
 \begin{align}\label{eq:concatenation_term3}
 \eqref{eq:last_term_convolution}&=\ampli\int_{0}^{T} dz_1 \int_{T}^{T+S} dz_4  K_0(x,L, z_1) K_0(L,z, T+S-z_4)  \itkernel( z_4-z_1).
 \end{align}
 Given that $\itkernel$ involves a Heaviside function, adding expressions \eqref{eq:concatenation_term1}, \eqref{eq:concatenation_term2} and \eqref{eq:concatenation_term3}, 
 one reproduces formula \eqref{eq:concatenation_summed}.

\section{Bounds involving the erfc function}\label{app:erfc}
There exist some simple upper bounds involving the complementary error function that serve to deal 
with the integrals appearing in the computations of this article.
First, consider the bounds
\begin{align}
 0\leq e^{a^2}\text{erfc}(a)\leq \frac{1}{\sqrt{\pi}a},
\end{align}
valid for $a>0$. Indeed, the lower bound is immediate, 
while from the definition of the erfc we have the following straightforward derivation
\begin{align}\label{eq:erfc_bound1}
 \begin{split}
\frac{2}{\sqrt{\pi}}e^{a^2}\int_a^{\infty} du \,e^{u^2}&=\frac{2a}{\sqrt{\pi}}\int_1^{\infty} du\, e^{-(u^2-1)a^2}\\
 &=\frac{2}{\sqrt{\pi}a} \int_0^{\infty} du\, e^{-u (\frac{u}{a^2}+2)}\\
 &\leq \frac{2}{\sqrt{\pi}a} \int_0^{\infty} du\, e^{-2u }\\
 &= \frac{1}{\sqrt{\pi}a}.  
 \end{split}
\end{align}

In addition, we can also prove that the propagator $\Kd$ has an upper bound for negative $\ampli$. 
The derivation, valid again for $a>0$, is {\it mutatis mutandis} the same as in expression \eqref{eq:erfc_bound1}:
\begin{align}\label{eq:erfc_bound2}
\begin{split}
 \frac{1}{\sqrt{\pi}a}-e^{a^2}\text{erfc}(a)&=\frac{1}{\sqrt{\pi}a}-\frac{2}{\sqrt{\pi}a} \int_0^{\infty} du e^{-u (\frac{u}{a^2}+2)}\\
 &=\frac{1}{\sqrt{\pi}a} \int_0^{\infty} du e^{-2u} \left(1-e^{-\frac{u^2}{a^2}}\right)\\
 &\leq\frac{1}{\sqrt{\pi}a} \int_0^{\infty} du e^{-2u} \frac{u^2}{a^2}\\
 &=\frac{1}{4\sqrt{\pi}a^3}. 
\end{split}
\end{align}

\section{Gel'fand-Yaglom approach}\label{app:GY}
In this appendix we obtain some of the previous results using the Gel'fand-Yaglom approach described in Ref.\cite{Fosco:2019lmw}. It has been shown there that the self-energy for a thin mirror 
with time-independent inhomogeneities reads
\begin{equation}
	E \;=\; \frac{1}{2} \,\int_{-\infty}^{+\infty} \frac{dk_0}{2\pi}
	{\rm Tr}  \log \Big[ {\mathbb I}  \,+\, \frac{1}{2\sqrt{{\mathbb
	H}_0}} \, \zeta(\paras ) \Big] \;,
\end{equation}
where ${\mathbb H}_0 = - {\mathbf\nabla}_\parallel^2 + k_0^2$. 
%From now on we omit the supraindex $\parallel$.

Writing
\begin{equation}
\zeta({\paras})=\zeta+\eta({\paras})\, ,
\end{equation}
we can expand the self-energy in powers of $\eta$. The second order reads
\begin{equation}
E^{(2)}=-\frac{1}{16}\int_{-\infty}^{+\infty} \frac{dk_0}{2\pi}{\rm Tr}\Big[\Big(
\frac{{\mathbb I} }{{\mathbb I} +\frac{\zeta}{2{\mathbb H}_0}}\, \frac{\eta}{{\mathbb H}_0}\Big)^2\Big]\, \label{E2}
\end{equation}
and corresponds to $\Gamma_{TI}^{(2)}/T_0$ in Eq.\eqref{eq:time_independent}.

The two-point function appearing in the above equation can be explicitly written as
\begin{equation}\label{N}
N({\paras}, {\paras[y]}):=
\frac{{\mathbb I} }{{\mathbb H}_0 +\frac{\zeta}{2}}\,  ({\paras}, {\paras[y]})=\int \frac{d\paras[k]}{(2\pi)^2}
\frac{e^{i\paras[k]\cdot(\paras - {\paras[y]})}}{\sqrt{k_0^2+{\paras[k]}^2}+\frac{\zeta}{2}}\, .
\end{equation}
Replacing Eq.\eqref{N} into Eq.\eqref{E2} we get
\begin{equation}
E^{(2)}=-\frac{1}{16} \int \frac{dk_0}{2\pi}\int  d\paras  \int d\paras[y]\,  N({\paras}, {\paras[y]})
\eta(\paras[y]) N({\paras[y]}, {\paras})\eta(\paras)
\end{equation}
In terms of Fourier transforms we find
\begin{equation}
E^{(2)}=-\frac{1}{2}\int\frac{d\paras[k]}{(2\pi)^2}\vert \tilde\eta(\paras[k])\vert^2
F(\paras[k],\ampli)\ , 
\end{equation}
with
\begin{equation}
F(\paras[k],\ampli)=\frac{1}{8}\int \frac{dp_0\,d\mathbf p}{(2\pi)^3}\frac{1}{\sqrt{p_0^2+(\mathbf p+\paras[k])^2}+\frac{\zeta}{2}}\, \frac{1}{\sqrt{p_0^2+\mathbf p^2}+\frac{\zeta}{2}}\, .
\end{equation}
This expression has an UV divergence, and some regularization is
implicitly assumed. However, we will compute
\begin{equation}
\Delta F(\paras[k],\ampli)=F(\paras[k],\ampli)-F(\mathbf 0,\ampli)\, ,
\end{equation}
which is finite and independent of the regulator.

To compute the integral we work in $p$-space spherical coordinates, and assume $\paras[k]=\vert \para[k]\vert \,\hat z$:
\begin{eqnarray}
\Delta F(\paras[k],\ampli)&=&\frac{1}{32\pi^2}\int_0^\infty dp \, p^2 \int_0^\pi d\theta\,  \sin\theta\,  G(p,\theta,\para[k])\\
G(p,\theta,\para[k])&=& \frac{1}{\frac{\zeta}{2}+\sqrt{p^2+{\para[k]}^2+2 p \para[k]\cos\theta}}\,\frac{1}{\frac{\zeta}{2}+p}-\frac{1}{(\frac{\zeta}{2}+p)^2}\, .
\end{eqnarray}
Performing the angular integral and defining as before $a= \zeta/\vert \para[k]\vert$ we find
\begin{eqnarray}
\Delta F(\para[k],\ampli)&=& \frac{\vert\para[k]\vert}{32\pi^2}\int_0^\infty dp\, p^2 \Big[ -\frac{8}{(a+2p)^2}+\frac{2}{p(2p+a)}\Big (1+p-\vert p-1\vert\nonumber\\ 
&+& \frac{a}{2}\log \Big(\frac{a+2\vert p-1\vert}{a+2(p+1)}\Big)\Big)\Big]
\end{eqnarray}
The integral in $p$ can be computed analytically, and the result
coincides, up to a constant term,  with the one obtained using the HK approach in Sec. \ref{sec:D4m0}. Indeed, we find
\begin{equation}\label{eq:deltaF}
\Delta F(\para[k],\ampli)=F_{NL}(\para[k],\ampli)-\frac{3\ampli}{32\pi^2}\, ,
\end{equation}
where $F_{NL}(\para[k],\ampli)$ is given in Eq.\eqref{eq:ffnl}.
Note that if we consider $\eta(\paras)$ as a background field, the constant term can be absorbed into a redefinition of its mass, as discussed in Sec. \ref{sec:D4m0}. Note also that, although in this appendix we considered time-independent inhomogeneities, on general grounds we expect
the form factor to be a function of the modulus of $\para[k]$, and therefore
the result in Eq.\eqref{eq:deltaF} is valid in the general case.

This calculation is a nontrivial cross-check of our
results, that also shows that the nonlocal part of the form factor does not depend on the regularization scheme.
\printbibliography
\end{document}